\documentclass[twocolumn]{aastex631}
\newcommand{\trackchange}[1]{{#1}}
\newcommand{\trackchanget}[1]{{#1}}

\usepackage[utf8]{inputenc}
\usepackage{tipa}
\usepackage{combelow}
\usepackage{breqn}
\usepackage{tabularx}

\shorttitle{JWST Quaoar Occultation}
\shortauthors{Proudfoot et al.}

\begin{document}

\title{Constraints on Quaoar's rings and atmosphere from JWST/NIRCam observations of a stellar occultation}

%\correspondingauthor{Benjamin Proudfoot}
%\email{benp175@gmail.com}

\author[0000-0002-1788-870X]{Benjamin C.N. Proudfoot}
\affiliation{Florida Space Institute, University of Central Florida, 12354 Research Parkway, Orlando, FL 32826, USA}

\author[0000-0002-6117-0164]{Bryan J. Holler}
\affiliation{Space Telescope Science Institute, Steven Muller Building, 3700 San Martin Drive, Baltimore, MD, 21218, USA}

\author[0000-0003-1260-9502]{Ko Arimatsu}
\affiliation{The Hakubi Center / Astronomical Observatory, Graduate School of Science, Kyoto University Kitashirakawa-oiwake-cho, Sakyo-ku, Kyoto 606-8502, Japan}

\author[0000-0002-6085-3182]{Flavia L. Rommel}
\affiliation{Florida Space Institute, University of Central Florida, 12354 Research Parkway, Orlando, FL 32826, USA}

\author[0009-0004-7149-5212]{Cameron Collyer}
\affiliation{Florida Space Institute, University of Central Florida, 12354 Research Parkway, Orlando, FL 32826, USA}

\author[0000-0003-2132-7769]{Estela Fern\'{a}ndez-Valenzuela}
\affiliation{Florida Space Institute, University of Central Florida, 12354 Research Parkway, Orlando, FL 32826, USA}

\begin{abstract} 

Observations of stellar occultations have revealed that small bodies are capable of hosting ring systems. The trans-Neptunian object (TNO) Quaoar, is the host of an enigmatic ring system, with two rings located well-outside the Roche limit. To better understand these structures, we observed a stellar occultation by Quaoar and its rings using the James Webb Space Telescope's (JWST) NIRCam instrument. Our observations detect both known rings, \trackchange{although Q2R---the inner known ring---is not detected on both sides of Quaoar, showing that it has substantial azimuthal variations similar to Q1R---the outer ring.} We also fit a model of the ring radii and pole orientation of the ring system, which confirms that Quaoar's spin-orbit and Weywot's mean motion resonances (especially Weywot's 6:1) may play a role in the rings' confinement and stability. In addition to examination of Quaoar's ring system, we also use our observations to place upper limits on a putative CH$_4$ atmosphere around Quaoar, finding that no global atmosphere with surface pressure \trackchange{$>1$} nbar can exist (at 3$\sigma$ significance). The lack of atmosphere supports the hypothesis that atmospheric processes are not the source of Quaoar's recently discovered inventory of light hydrocarbons.

\end{abstract}
\keywords{Planetary rings (1254), Stellar occultation (2135), Trans-Neptunian objects (1705), James Webb Space Telescope (2291)}

\section{Introduction}
\label{sec:intro}

In the past decade, rings have been discovered around small solar system bodies \citep{braga2014ring,ortiz2017size,morgado2023dense,pereira2023two,ortiz2023changing}, marking a significant advancement in planetary science. Their discoveries---enabled through observing stellar occultations by small bodies---have permitted a new way to study the history and evolution of small bodies. Thus far, four ring systems have been found, two around Centaurs and two around the large trans-Neptunian objects (TNOs) Haumea and Quaoar. The formation, evolution, and long-term stability of small body rings are hotly debated \citep[e.g.,][]{morgado2023dense,winter2019location,marzari2020ring,madeira2022dynamics,rodriguez2023dynamical}. As many small bodies have significantly non-axisymmetric shapes, the dynamics of small body rings are not just subject to satellite-ring interactions (like those around the giant planets) but are also subject to interactions with the shapes of their hosts \citep[see][for a detailed review]{sicardy2020dynamics}. Understanding of small body rings is still in its infancy, although it will inevitably draw on a strong foundation developed over decades of studying giant planet rings. 

The TNO Quaoar is the host of the most mysterious small body ring system, composed of at least two rings that were revealed by stellar occultations \citep{morgado2023dense,pereira2023two}. These rings were found well outside Quaoar's Roche limit, even accounting for the low densities of possible ring material. The outer ring---referred to as Q1R---is relatively wide ($>100$ km) with shallow optical depth, except during a few detections where the ring was dense and thin ($\sim5-10$ km). These variations in width and depth have been interpreted as dense arcs embedded within a continuous ring. The inner ring, Q2R, has only been detected during a single event (although from multiple telescopes), and little is known about it. Explaining Quaoar's rings has proved difficult, as rings outside of a body's Roche limit are expected to accrete into moonlets on decades-long timescales \citep{Kokubo2000,Takeda200}. One explanation for the rings' presence outside of the Roche limit posits that the ultra-cold temperature of ring particles at Quaoar's heliocentric distance may promote more elastic collisions, preventing accretion. However, this mechanism causes viscous spreading---which has to be balanced by other mechanisms---and is much less effective in the denser portions of the ring \citep{morgado2023dense}. 

The confinement of both rings (especially if viscous spreading is rapid) is likely provided by both spin-orbit resonances (SORs), between the rings and Quaoar's non-axisymmetric gravitational field, and mean motion resonances (MMRs) with Quaoar's moon Weywot. The outer, denser ring is near both Quaoar's 3:1 SOR\footnote{Technically SORs around triaxial bodies with the form $n$:$m$ must have $m$ as a multiple of two, so only the 6:2 SOR technically exists \citep{1999ssd..book.....M,sicardy2020resonances}. For brevity and to ease comparison with other work, we will use the reduced version.}
and Weywot's 6:1 MMR \citep{morgado2023dense}. Likewise, the inner ring is near Quaoar's 7:5 resonance \citep{pereira2023two}. These resonances may contribute to the confinement of ring particles, and may also serve as mechanisms to further excite collisions. Unfortunately, current understanding of the rings' interactions with both SORs and MMRs is severely hampered by Quaoar's uncertain shape (compare \citealp{braga2013size} and \citealp{kiss2024visible}) and lack of a high-quality solution for Weywot's orbit \citep{vachier2012determination,morgado2023dense}.

On the other hand, it may be possible for the rings to be relatively young and constantly replenished by delivery of fresh material. Recent spectral observations of Quaoar by JWST's NIRSpec integral field unit (IFU) have revealed the presence of CH$_4$ on Quaoar's surface which, given its relatively short surface lifetime (due to both Jeans escape and irradiation), may imply ongoing resupply to the surface \citep{emery2024tale}. Periodic outgassing or geological activity could periodically refresh Quaoar's surface CH$_4$ inventory while delivering new material to its rings \citep{Anand2023Origin}. \trackchange{Another potential source could be from impacts on Quaoar, which could excavate material that ends up on stable ring-like orbits.} Although speculative, there is significant uncertainty around the rings' long-term stability---especially with respect to the relatively unexplored dynamical environment---which motivates exploration of all hypotheses.

%Currently, the only way to study Quaoar's rings is through the observation of stellar occultations. One of the best facilities to observe occultations is the James Webb Space Telescope (JWST), due to its large aperture, photometric stability, and rapid readout. Although JWST provides just one chord, the results from such a chord are valuable for furthering our understanding of the physical processes at work around Quaoar. 

In this work, we present JWST observations of a recent stellar occultation by Quaoar and its rings. We first present the details of the occultation prediction, observations, and data analysis (Section \ref{sec:methods}). Then, we show the detections of structures surrounding Quaoar (Section \ref{sec:detect}), which are combined with all past detections to construct a geometric model of the rings (Section \ref{sec:model}). Next, we derive new constraints on Quaoar's atmosphere from the non-detection of atmospheric refraction near occultation (Section \ref{sec:atmosphere}). Lastly, we discuss the implications of our findings (Section \ref{sec:discussion}) and conclude (Section \ref{sec:conclusion}). 

\section{Prediction, Observations, and Data Analysis}
\label{sec:methods}

\begin{figure}[ht]
    \centering\includegraphics[width=0.99\columnwidth]{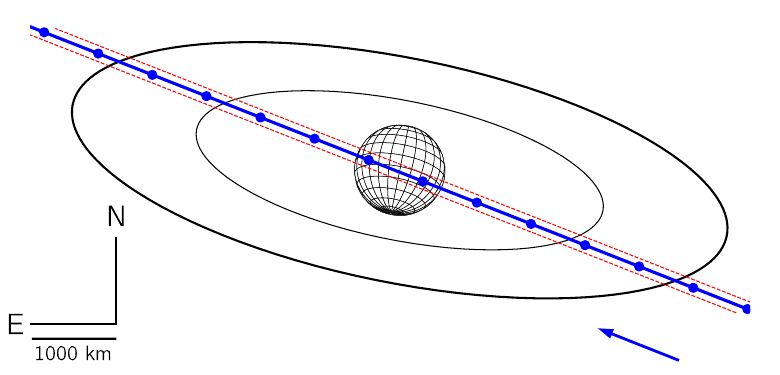}
    \caption{Predicted path of the occulted star through the Quaoar system as seen from JWST (in blue), with $3\sigma$ ephemeris uncertainties (dashed red lines). The prediction was based on the ephemerides of Quaoar and JWST from JPL and NAIF, respectively (see text for more details of the ephemerides). Dots correspond to the predicted position every minute. Quaoar itself is modeled as a sphere with a radius of 550 km. The rings are modeled with radii and pole position as given in \citet{pereira2023two}. }
    \label{fig:pred}
\end{figure}

Using the ephemerides of Quaoar and JWST from JPL Horizons\footnote{\url{https://ssd.jpl.nasa.gov/horizons/}} and the Navigation and Ancillary Information Facility\footnote{\url{https://naif.jpl.nasa.gov/pub/naif/JWST/kernels/spk/}}, respectively, we used the Stellar Occultation Reduction and Analysis tool \citep[SORA;][]{Gomes-Junior2022} to predict that Quaoar would occult a star within the Gaia catalogue on 2024-08-28 UT. 
%As Quaoar was near quadrature during the event, the on-sky velocity was only 11.78 km s$^{-1}$, allowing for better spatial resolution than typical occultation events ($\sim$20 km s$^{-1}$). 
The occulted star (Gaia DR3 4103291293710742656, \trackchange{J2000 ICRS position $\alpha = 18^h31^m31.2286^s$, $\delta = -15\degr07'52.25''$}) has a Gaia $G$-band magnitude of 17.99 mag, $\sim$3-5 times brighter than Quaoar (depending on the wavelengths observed). \trackchange{Relative velocity of the star with respect to Quaoar was 11.78 km s$^{-1}$.}
A schematic showing the predicted path of the star through the Quaoar system is shown in Figure \ref{fig:pred}. 

As discussed in \cite{santos2016occult}, predictions of occultations visible from JWST are only valid for a few months prior to the event. This is due to JWST's unstable orbit around the L2 point, requiring station-keeping maneuvers every $\sim$6 weeks \citep{Gardner2023}. This creates difficulty for planning observations within JWST's yearly proposal cycle. As such, only target of opportunity or Director's Discretionary time (DDT) proposals can effectively allow for JWST observations of stellar occultations by solar system objects. The observations presented here were made as part of DDT program 6780.

Observations were acquired using the NIRCam \trackchange{\citep{rieke2023performance}} time series imaging mode at an imaging cadence of $\sim$5 Hz, chosen as a compromise between spatial resolution of the occultation chord and signal-to-noise ratio (SNR) on the occulted star. Effective exposure time for each individual integration was 0.15 s. NIRCam allows simultaneous imaging in both the short and long wavelength channels (SWC and LWC, respectively) with the use of a dichroic beam splitter. In order to maximize the SNR, we chose the widest possible filters, F150W2 \trackchange{(with wavelength from 1.01--2.38 $\mu$m)} and F322W2 \trackchange{(with wavelength from 2.43--4.01 $\mu$m)}, in the SWC and LWC, respectively.

The raw {\it uncal} time-series data for each filter were downloaded from the Mikulski Archive for Space Telescopes (MAST), which can be found at: \dataset[10.17909/g07f-g874]{http://dx.doi.org/10.17909/g07f-g874}. A total of 26,575 integrations were obtained using the SUB64P subarray, the RAPID readout pattern, and 3 groups per integration. The separate files for the F150W2 and F322W2 filters were first broken into individual integrations (samples) and timestamped, then processed locally into flux-calibrated {\it cal} files using JWST calibration pipeline version 1.15.1 and the {\it jwst\_1274.pmap} context file \citep{Bushouse2024}. The reason for breaking up the time series for processing was twofold: (1) \trackchange{to remove the vertical banding pattern noise} from intermediate {\it rate} images using the routine developed by Dan Coe\footnote{\url{https://github.com/dancoe/NIRCam/blob/main/NIRCam\%20smooth1overf.ipynb}} and (2) prevent the {\it jump} step of the pipeline from flagging and removing thousands of samples around the Quaoar solid-body occultation. The {\it jump} step not only  identifies discontinuous jumps in flux within an integration due to a cosmic ray strike, it also identifies increases in flux within a given pixel between subsequent integrations. Due to the increase in flux as Quaoar approached the star, the entirety of the Quaoar solid-body occultation, including immersion and emersion, was flagged and removed by the pipeline. This pipeline behavior, which could not be turned off via optional arguments, was avoided by processing the integrations separately.

Following calibration, the centroid of the occulted star was calculated from the average of the first 1000 images taken in each filter. This was used as an initial guess, with the centroid then more precisely determined in each individual {\it cal} image and used for centering the circular extraction aperture. Extraction aperture radii were 1.5 times the full-width at half-maximum (FWHM) of the point-spread function in each filter, which is 1.452 and 1.524 pixels for the F150W2 and F322W2 filters, respectively. Inner and outer background radii were 4$\times$FWHM and 7$\times$FWHM, respectively. The background was removed by computing the median within the annulus and subtracting that value from every pixel in the image. The median absolute deviation within the annulus was computed and added in quadrature with the pipeline-computed errors. The {\it photutils} package was used to perform the photometry on the background-subtracted images and to calculate the uncertainties \citep{Bradley2024}. Finally, the light curve in each filter was normalized by the median of the first 10,000 measurements (i.e., baseline measurements outside of the ring or solid-body occultations).

\begin{figure*}
    \centering\includegraphics[width=0.99\textwidth]{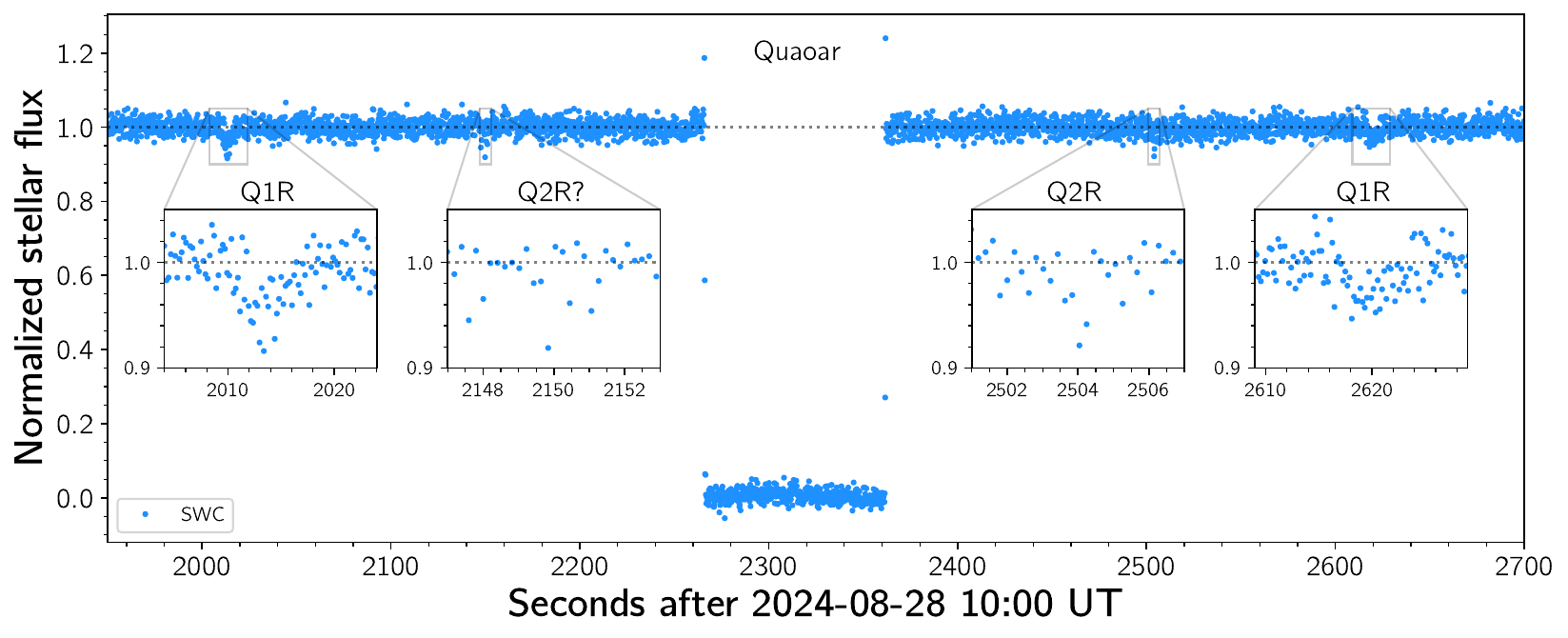}
    \caption{The normalized occultation light curve in the short wavelength channel (SWC). \trackchange{Photometric uncertainties are 1.8\%.} Labeled insets show the four ring/arc detections in the light curve. We note that V-shapes seem to be present in both detections of Q1R.}
    \label{fig:lc}
\end{figure*}

\begin{figure*}
    \centering\includegraphics[width=0.99\textwidth]{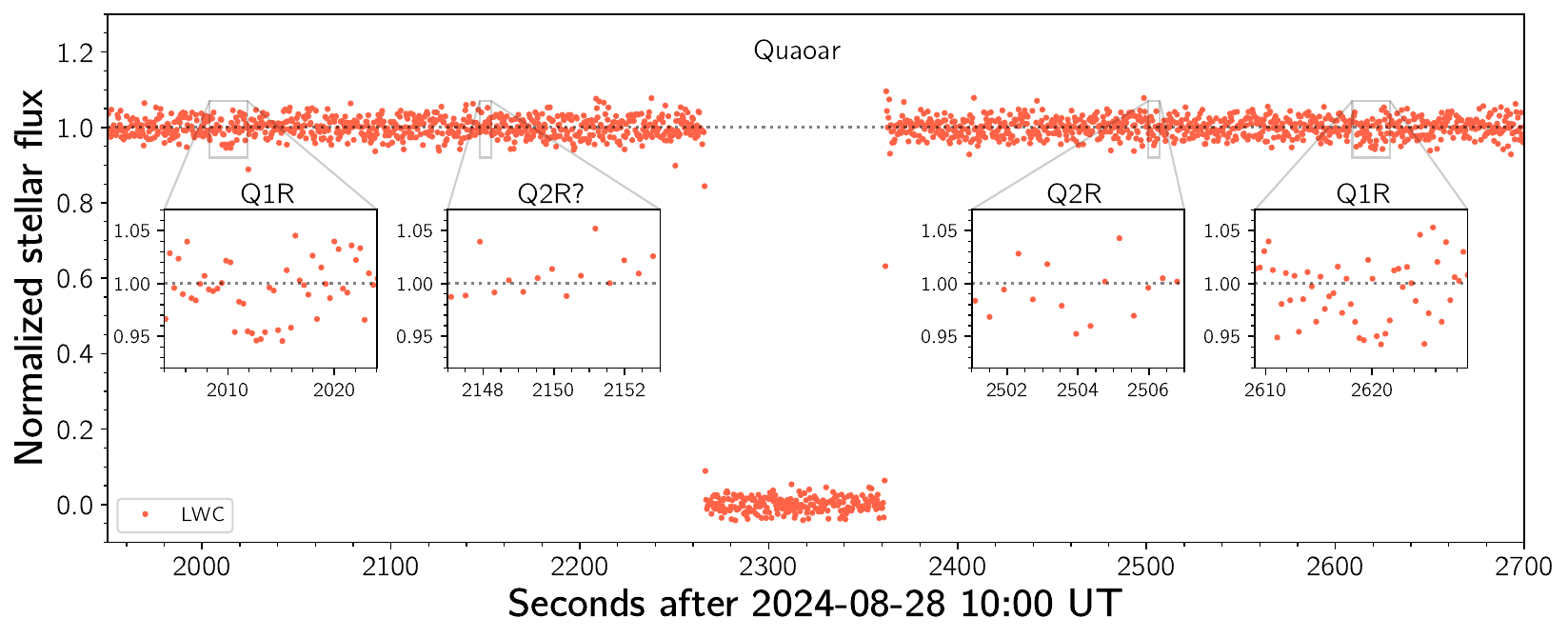}
    \caption{\trackchange{The normalized occultation light curve in the long wavelength channel (LWC) in the style of Figure \ref{fig:lc}. Photometric uncertainties are 2.5\%. Data is binned by two, which acts to smooth the data and increase SNR.}}
    \label{fig:lc_lw}
\end{figure*}

%\efv{We performed synthetic aperture photometry on the target star using our own routines written in Python. As Quaoar moves toward the star, the PSFs of both Quaoar and the star merge together, causing the centroid calculation to shift instead of remaining steady at the star's photocenter. To mitigate this, we calculated the centroid of the star using a set of images acquired at the beginning of the observation, before the PSF of Quaoar began to blend with that of the star. The aperture radius was selected to maximize the SNR of the star, thereby minimizing photometric dispersion ($\sigma = xx$ and $xx$ mag for the F150W2 and F322W2 filters, respectively).}

\trackchange{Quaoar was $\sim$35\% and $\sim$15\% as bright as the occulted star in the SWC and LWC, respectively, and was clearly detectable in individual images.} Near the time of occultation, Quaoar's entry into the \trackchange{photometric} aperture slowly increased the total flux measured, \trackchange{producing a distinct} ``hump''-like signal centered on the occultation mid-time. 
%Even inside of occultation when the star's flux has been completely blocked, the hump trend continues. 
Removal of this structure is necessary for proper detection and characterization of the surrounding ring system. \trackchange{We removed Quaoar's constant (over the short timescale) flux contribution, using a} Savitzky-Golay (SG) digital filter, as implemented in SciPy \citep{virtanen2020scipy}. We normalized the data using a fourth-degree polynomial fit to each batch of 2001 images. We note that a different method, which implements a high-pass filter to remove any unwanted low-frequency signals, produces a nearly identical light curve. Much of our analysis was run on both \trackchange{the SG filtered and high-pass filtered light curves}, showing very similar results. For brevity, we choose to present the light curve/results from the SG filtering method. Our normalized light \trackchange{curves are shown in Figures \ref{fig:lc} and \ref{fig:lc_lw} which have photometric uncertainties of 1.75\% and 2.5\% in the SWC and LWC, respectively}.

The timings and uncertainties of immersion and emersion from occultation (both the solid body and the rings) were obtained with the SORA package, which uses $\chi^2$-minimization to obtain the light curve model that best represents the observed data \citep{Gomes-Junior2022}. \trackchange{Rings were modeled as a grey screen with an apparent opacity as a fitted parameter in the $\chi^2$-minimization.} The model considers the exposure time, the Fresnel scale, and star diameter effects. 

The Fresnel scale fringe effects are given by
\begin{equation}
    F = \sqrt{\frac{\lambda \Delta}{2}}
\end{equation}
where $\lambda$ is the observational wavelength and $\Delta$ is Quaoar's distance from JWST's location at the time of the event (42.12 au). We used the central wavelength of each filter, 1.6865 $\mu$m and 3.2244 $\mu$m to estimate a Fresnel scale of 2.3 km and 3.2 km for the SWC and LWC light curves, respectively.

\trackchange{The stellar apparent angular diameter was calculated by using the \citet{Kervella2004} and \citet{vanbelle1999} formalism. However, since the star has no information about its Johnson-Cousins V and K magnitudes, we used the \textit{Gaia} catalog G, BP, and RP magnitudes to obtain them by using the photometric relation provided by \citealt{Riello2021} (see Table C.2). The obtained stellar angular diameter at Quaoar's geocentric distance (42.12 au) ranges from 0.0051 mas (0.155 km) for a main sequence star to 0.0068 mas (0.207 km) if we consider it a variable star. Hence, the occultation light curve is dominated by the exposure time (1.77 km) and Fresnel effects (3.8 km).}

% We estimate the stellar angular diameter at Quaoar's distance using the method of \citet{Kervella2004}. Since the target star has no information about its Johnson-Cousins magnitudes, we converted \textit{Gaia} catalog magnitudes to the Johnson-Cousins photometric system (see Table C.2 in \citealt{Riello2021}) to estimate a stellar angular diameter $\theta_V$ = 0.18 km or 0.0058 mas. Hence, the light curve is dominated by the exposure time and Fresnel effects, both of which are the same order of magnitude. 

The results of our light curve modeling are shown in Table \ref{tab:detect}. 

\begin{deluxetable*}{lcccccccc}
\tablecaption{Quaoar Occultation Circumstances}
\tablehead{
Ring        &  & Immersion                      & Emersion                       & $W_r$      & $p_n$               & $\tau_n$             & $E_p$ & $A_\tau$ \\
            &           & 2025-08-28                     & 2025-08-28                     & (km)       &                 &                 & (km)             & (km)
}
\startdata
Q1R ing. & SWC       & 10:33:31.901$\pm$0.053           & 10:33:36.197$\pm$0.077           & 50.6$\pm$1.1 & 0.0062$\pm$0.0006 & 0.0062$\pm$0.0006 & 0.31$\pm$0.03      & 0.31$\pm$0.03      \\
            & LWC       & 10:33:30.509$\pm$0.211           & 10:33:36.131$\pm$0.076           & 66.3$\pm$2.6 & 0.0056$\pm$0.0009 & 0.0056$\pm$0.0009 & 0.37$\pm$0.06      & 0.37$\pm$0.06      \\
\hline
Q1R egr.  & SWC       & 10:43:37.451$\pm$0.404           & 10:43:43.542$\pm$0.117           & 71.8$\pm$5.0 & 0.0040$\pm$0.0006 & 0.0040$\pm$0.0006 & 0.29$\pm$0.05      & 0.29$\pm$0.05      \\
            & LWC       & 10:43:38.064$\pm$0.290           & 10:43:41.704$\pm$0.116           & 42.9$\pm$3.7 & 0.0056$\pm$0.0012 & 0.0056$\pm$0.0013 & 0.24$\pm$0.06      & 0.24$\pm$0.06      \\
\hline
Q2R ing. & SWC       & 10:35:49.693$\pm$0.082           & 10:35:49.903$\pm$0.060           & 2.5$\pm$1.2  & 0.0226$\pm$0.0130 & 0.0234$\pm$0.0140 & 0.06$\pm$0.04      & 0.06$\pm$0.05      \\
            & LWC       & \nodata                        & \nodata                        & \nodata    & $\lesssim$0.005         & $\lesssim$0.005         & \nodata          & \nodata          \\
\hline
Q2R egr.  & SWC       & 10:41:43.715$\pm$0.068           & 10:41:44.293$\pm$0.060           & 6.8$\pm$1.1  & 0.0105$\pm$0.0019 & 0.0107$\pm$0.0019 & 0.07$\pm$0.02      & 0.07$\pm$0.02      \\
            & LWC       & 10:41:43.547$\pm$0.203           & 10:41:44.507$\pm$0.244           & 11.3$\pm$3.7 & 0.0084$\pm$0.0031 & 0.0084$\pm$0.0032 & 0.10$\pm$0.05      & 0.10$\pm$0.05     \\
\hline
Quaoar & SWC & 10:37:46.223$\pm$0.002 & 10:39:21.656$\pm$0.004 & 1124.68$\pm$0.05 & \nodata & \nodata & \nodata & \nodata \\
       & LWC & 10:37:46.229$\pm$0.011 & 10:39:21.671$\pm$0.003 & 1124.78$\pm$0.13 & \nodata & \nodata & \nodata & \nodata
\enddata
\tablecomments{All values are calculated based on our occultation modeling as described in Section \ref{sec:methods}. The bottom two rows show the immersion, emersion, and chord length of the Quaoar solid body occultation. Normal opacity ($p_n$) and normal optical depth \trackchanget{($\tau_n$)---as defined in \citet{morgado2023dense} and \citet{pereira2023two}---}are calculated based on the ring opening angle from JWST \trackchange{($18.1\pm0.2\degr$)} and the ring pole orientation found in Section \ref{sec:model}. This corrects for the geometric effects of viewing the rings edge on. Equivalent width and equivalent depth are defined as $E_p = p_nW_r$ and $A_\tau = \tau_nW_r$, assuming the square-well model.}
\label{tab:detect}
\end{deluxetable*}

\section{Detection of Rings Around Quaoar}
\label{sec:detect}

\begin{figure}
    \centering\includegraphics[width=0.49\textwidth]{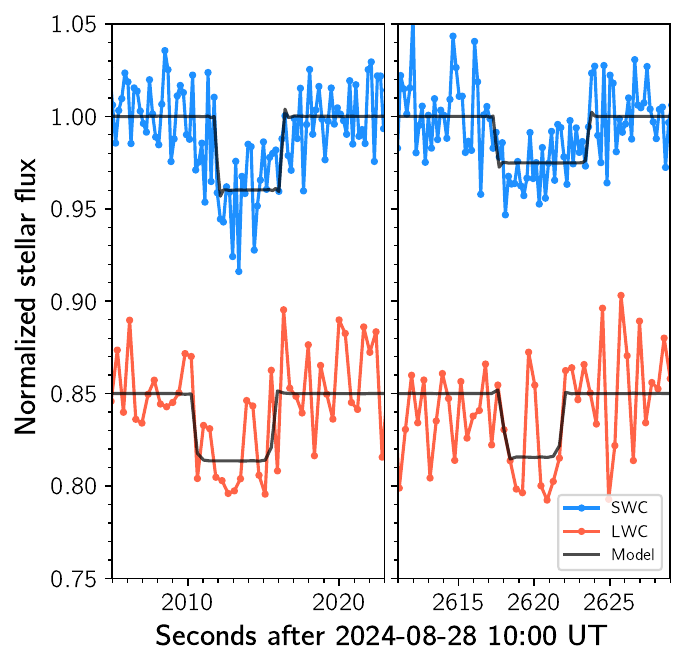}
    \caption{Detections of Q1R on ingress (left) and egress (right). In blue and red are the occultation light curve in NIRCam's SWC and LWC, respectively. In black are square-well model fits convolved with star diameter and Fresnel diffraction effects. Both filters show statistically significant detections of Q1R. }
    \label{fig:q1r}
\end{figure}

When examining the normalized light curve (Figure \ref{fig:lc}), Q1R can be easily identified by-eye. In Figure \ref{fig:q1r}, we show our F150W2 and F322W2 detections of Q1R alongside our model of the light curve. Full details of the measured opacity, ring width, and other model parameters are given in Table \ref{tab:detect}. For Q1R, for which we have detections on both ingress and egress (i.e., before and after the main body)\footnote{For the rest of this paper, we use ingress and egress to refer to detections before and after the main body. This in contrast to immersion and emersion, which we use to mean disappearance and reappearance of the star.}, we found no variable width and opacity, unlike previous detections \citep{morgado2023dense,pereira2023two}. We attribute this symmetry to a chance alignment with the orbiting density variations within Q1R.

Interestingly, in both the SWC and LWC, hints of structure appear within Q1R. In the SWC, there is a V-shaped profile on both ingress and egress, which is easily visible in Figure \ref{fig:lc}. Given the noisy data, it is possible that these shapes are due to random chance, as indicated by the good fit quality of the square-well model \trackchange{(with $\chi^2$ per degree of freedom of 0.998 and 1.035 on ingress and egress, respectively)}, but the symmetry of the V-shaped profile potentially indicates the structure may be real. The LWC light curve also exhibits some structure, showing a clear W-shaped profile on egress, alongside a possible W-shape on ingress. Especially for the LWC, the data are noisy and the shape could be due to random chance, but the similarity of the structure in both LWC detections is suggestive. Unfortunately, given the data quality, these possible structures can only hint at the presence of structure within the rings. Structure within rings is common and can be the result of various macro- and micro-physical particle interactions \citep[e.g. ][]{brophy1990phase,meinke2012classification}.

\begin{figure}
    \centering\includegraphics[width=0.49\textwidth]{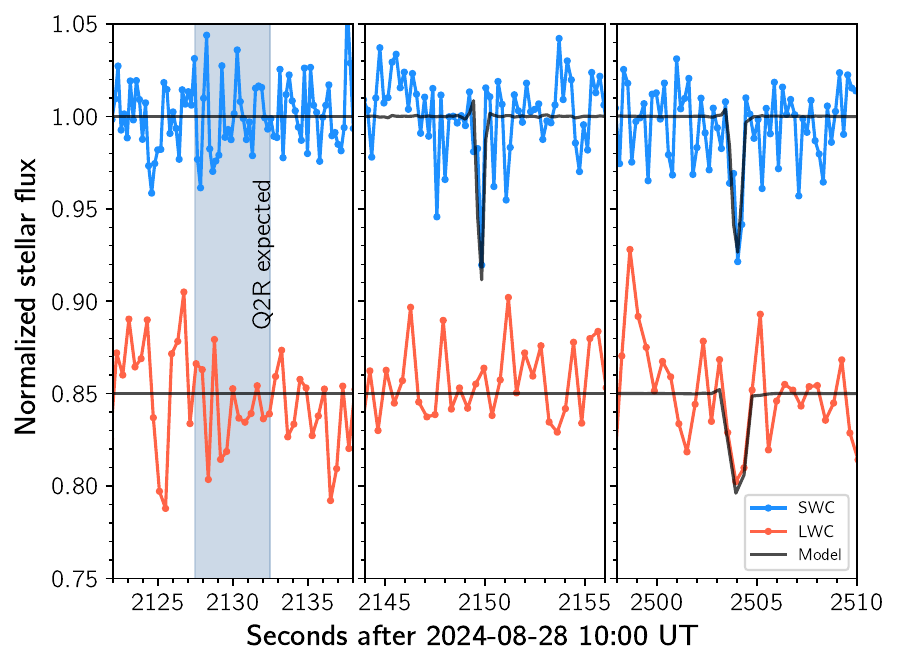}
    \caption{Detections of Q2R in the style of Figure \ref{fig:q1r}, with the expected location of Q2R on ingress (left), the possible detection inside of Q2R (center), and on egress (right) at the expected position. The grey region shows the expected location of Q2R on ingress, where there is notably no detection. }
    \label{fig:q2r}
\end{figure}

Our ring detections for Q2R are shown in Figure \ref{fig:q2r}. This marks the first published recovery of Q2R since the ring was discovered. Unlike for the denser Q1R, we only clearly detect Q2R at its expected position during egress. On ingress, some possible hints of variation exist, but are inconclusive. An upper limit on the normal \trackchanget{opacity} of Q2R on ingress varies strongly with the putative width of the ring. Assuming a thin ring of $\sim5$ km, we place an upper limit on the normal \trackchanget{opacity} on ingress of $p_n\lesssim$0.005 (2$\sigma$). The faintest detection of Q2R in the literature is near this detectability limit \citep{pereira2023two}, so a cut of Q2R that is slightly less dense (or less wide) could have gone undetected. Comparing our egress detection shows a substantially denser ($\sim2$-$5\times$) ring than other Q2R detections, confirming that Q2R has azimuthal variations. These are not as substantial as those in Q1R, but it clearly shows that Q2R is not uniform.

\trackchange{As can be clearly seen in Figures \ref{fig:lc_lw}, \ref{fig:q1r}, and \ref{fig:q2r}, the quality of the LWC light curve leaves much to be desired. Q1R is detected in the LWC at $\sim$6$\sigma$ and $\sim$4$\sigma$ confidence on ingress and egress, respectively. Based on these confidence levels, combined with the simultaneous detections in the SWC, these detections are robust. For Q2R in the LWC, our egress detection is at $\sim$2 $\sigma$ confidence. As can be seen in the LWC light curve, various other drops are near the same depth, casting some doubt on our claimed detection. However, just like with Q1R, the simultaneous drop in the SWC indicates that the detection is indeed robust, although any inferences based on this detection should be made cautiously.}

\trackchange{Comparing the optical depth of the SWC and LWC ring detections can place constraints on the composition and size of ring particles. Since the LWC was low SNR, relatively little can be authoritatively stated about the relative ``colors'' of either ring, especially so for Q2R. On both ingress and egress, the Q1R and Q2R opacity/optical depths are consistent between the SWC and LWC within $\sim$1 sigma. Although detailed characterization requires radiative transfer modeling, this suggests that size of the dominant grains/particles are larger than 3-4 $\mu$m, as smaller particles would less effectively absorb $\sim$3 $\mu$m light. Unfortunately, given the low quality of the LWC light curve, we leave full radiative transfer modeling to other events with higher SNR. Future multi-wavelength studies of Quaoar's rings should focus on high-SNR in both filters to allow for robust study.}

Approximately 20 seconds after the expected time of a dropout from Q2R on ingress, the SWC light curve exhibits a $>4\sigma$ drop in flux in a single image without a corresponding drop in the LWC. Similar drops were not seen at the same radial location on egress. Manual inspection of the image itself showed no obvious issues or defects. Likewise, both data normalization schemes show the drop at high significance. This deep drop could be explained by random chance; although a $4\sigma$ event is unlikely to occur at any given point, we would expect to see a $4\sigma$ outlier one time given the number of images acquired. Given the low probability of such an event combined with its occurrence within Quaoar's ring system, further discussion of this possible detection seems warranted. 

\begin{figure*}
    \centering\includegraphics[width=0.8\textwidth]{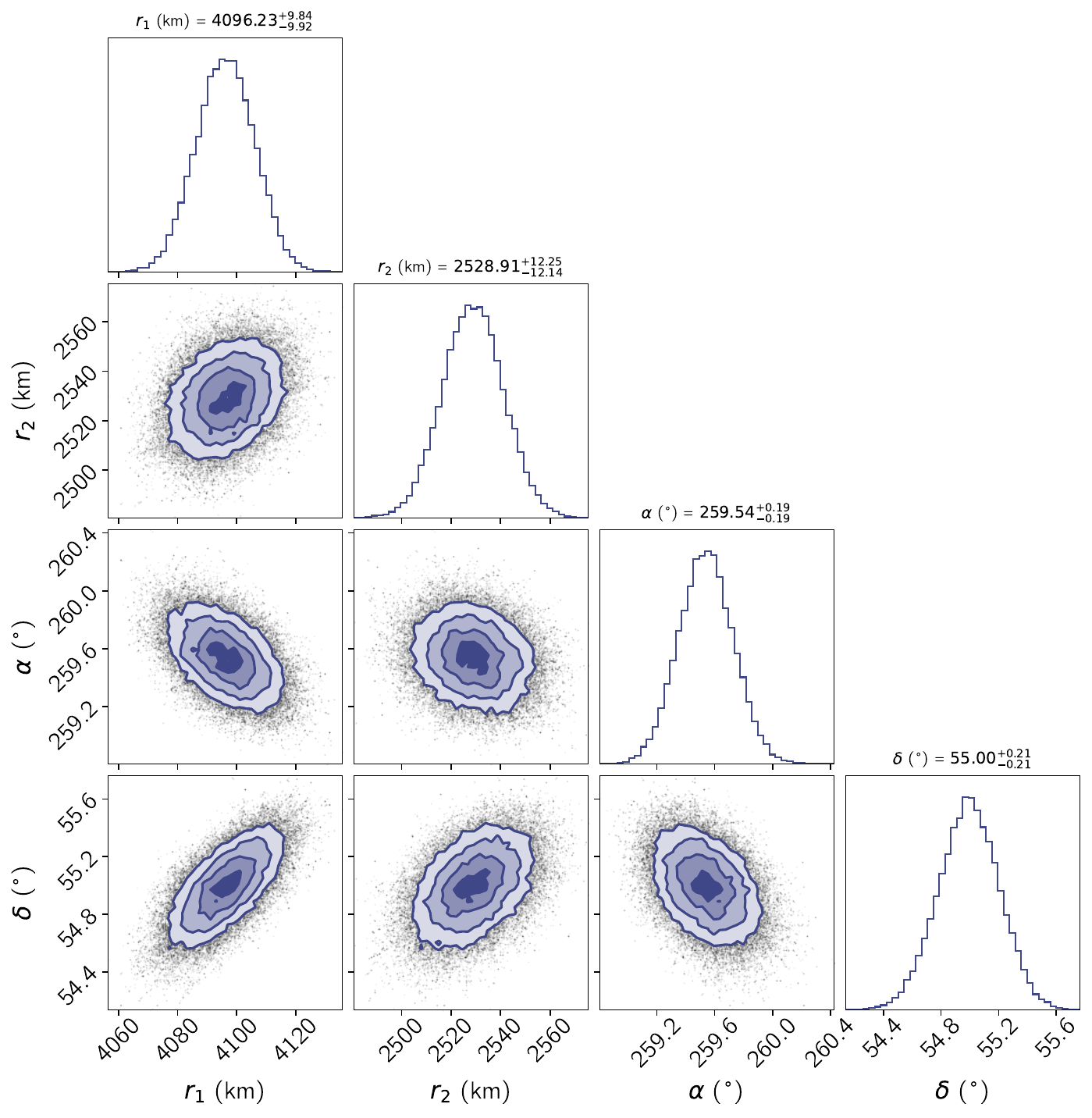}
    \caption{A corner plot showing the results of the ring modeling. Joint (2-dimensional) posterior distributions are shown as contour plots and marginal (1-dimensional) posteriors are shown as histograms at the top of each column.}
    \label{fig:corner}
\end{figure*}

Leaving aside the possibility of a random outlier\trackchange{---which seems likely to be the most parsimonious explanation---}the detection can be explained by two possibilities. Firstly, it could be explained by another narrow, diffuse ring around Quaoar\trackchange{, which remains undetected on egress and in the LWC light curve.} Alternatively, the unexpected drop may actually be Q2R, which would then imply substantial eccentricity and/or inclination relative to the circular Q1R. \trackchange{Both of these explanations seem unlikely, but} are easily testable. We discuss \trackchange{the evidence for or against them} further in Section \ref{sec:discussion}\trackchange{, showing that the outlier hypothesis seems to be the best explanation}.

\begin{figure*}
    \centering\includegraphics[width=0.8\textwidth]{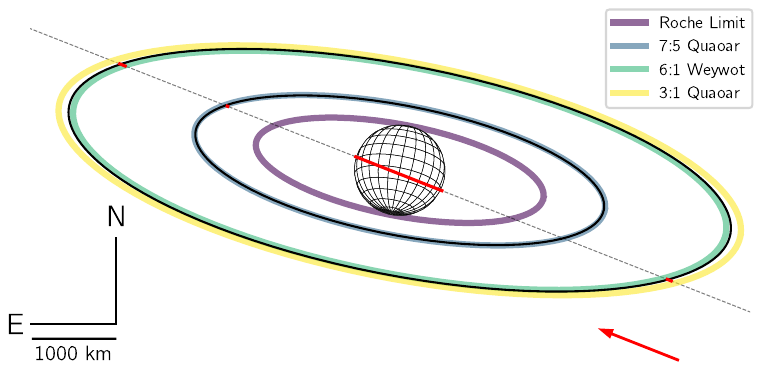}
    \caption{The three confirmed ring detections (in red) overlaid on a model of the Quaoar system. Colored ellipses in the legend are labeled from inside out. Ring parameters (radius and pole orientation) are taken from the results of our model fits in Section \ref{sec:model} \trackchanget{and the radius for the Roche limit taken from \citet{pereira2023two}}. Since our single chord provides little shape information about Quaoar itself, we show Quaoar as a sphere with a radius of 550 km. Note that the ring widths are not to scale in this schematic.}
    \label{fig:detect}
\end{figure*}

\section{The Geometry of Quaoar's Rings}
\label{sec:model}

With over two dozen detections of rings (of both Q1R and Q2R) the geometry of Quaoar's rings can be precisely measured. To do this, we combine all detections of the rings published in the literature and the detections from this work and simultaneously fit a model of both rings. Here, we assume both Q1R and Q2R have the same, fixed ring pole unless otherwise stated. At first, we discuss the results of the ring fitting without the possible ingress detection of Q2R. After, we discuss fits of Q2R with the ingress detection.

Since the viewing geometry slowly changes over time and observing location, we use a fitting approach where we reproject the on-sky coordinates of each ring detection ($f$, $g$ \trackchange{defined by their position with respect to the NIMAv19\footnote{\url{https://lesia.obspm.fr/lucky-star/obj.php?p=1177}} ephemeris of Quaoar, \citet{desmars2015orbit}}) onto an arbitrary ring plane (defined by its pole direction, $\alpha$, $\delta$\trackchange{, in J2000 ICRS coordinates}). The transformed coordinates ($x$, $y$) are then compared to a model of circular rings (with radii $r_1$ and $r_2$), by calculating the $\chi^2$. We define the $\chi^2$ as:
\begin{equation}
    \chi^2 = \sum^{i} \frac{\left(\sqrt{x_i^2+y_i^2} - r_1\right)^2}{\sigma^2_{i} + \sigma_e^2} + \sum^{j} \frac{\left(\sqrt{x_j^2+y_j^2} - r_2\right)^2}{\sigma^2_{j} + \sigma_e^2}
\end{equation}
\noindent where $x_{i}, y_i$ ($x_{j}, y_j$) is the reprojected coordinates of the $i$th ($j$th) detection of Q1R (Q2R) from the ephemeris center, $\sigma_{i}$ \trackchanget{and $\sigma_{j}$ are} the radial uncertainty of each detection, and $\sigma_e$ is the uncertainty in the ephemeris center. To match past work, we use $\sigma_e = 27$ km, which translates to about a 1 mas uncertainty in the ephemeris \citep[][]{morgado2023dense}. We also make use of the NIMA ephemeris of Quaoar, to maintain consistency with past work \citep{desmars2015orbit}. 

Using Bayesian statistics, the calculated $\chi^2$ of an arbitrary ring model (with parameters $r_1$, $r_2$, $\alpha$, and $\delta$), is then equivalent to $-2 \times \ln{\mathcal{L}(f,g|r_1,r_2,\alpha,\delta)}$, where $\mathcal{L}(f,g|r_1,r_2,\alpha,\delta)$ is the likelihood of the data given a circular, fixed-pole ring model. We can then explore the posterior probability ($\mathcal{L}(r_1,r_2,\alpha,\delta|f,g)$) using our likelihood function in a Bayesian parameter inference framework. Specifically, we use \texttt{emcee} \citep{foreman2013emcee}, a Markov Chain Monte Carlo (MCMC) ensemble sampler, to explore the parameter space of the model (given the observed data). In addition to our likelihood function defined above, we use uninformative priors on the radius of the ring and use a spherically uniform prior on the ring pole direction. In practice, this results in a uniformly distributed prior on $\alpha$ and a prior on $\delta$ that is proportional to $\cos(\delta)$. We ran the ensemble sampler with 100 walkers for a 500 step burn-in. After discarding the burn-in, we ran the sampler for 1000 steps. Convergence was confirmed based on examination of trace plots. The best-fit solution found with our sampler had a reduced $\chi^2_\nu = 0.92$, \trackchange{where $\chi^2_\nu$ is the $\chi^2$ per degree of freedom,} indicating a good fit to the data.

The results of our ring modeling are shown in Figure \ref{fig:corner} and Table \ref{tab:fits}. We also show a schematic of the system, as seen from JWST, with our detections and favored ring geometry in Figure \ref{fig:detect}. We note that our solution is somewhat different than those in the literature. Formally, our solution disagrees with the literature models at $>5\sigma$ significance, with solutions discrepent by 1.7$\degr$. 
\trackchange{This discrepancy appears to be systematic, as a fit excluding our detections does not reproduce the literature values. The source of this systematic error is unknown, but is likely related to exact modeling of the changing viewing angle. We do point out that} our solution has a $\chi^2_\nu$ lower than other solutions \citep{pereira2023two} and visualization of our detections in $f-g$ space (like in Figure \ref{fig:detect}) shows \trackchange{the literature ring pole solutions are clearly inconsistent with our detections}.
%In the past, Q1R detections were fit without accounting for the changing viewing geometry. Although this greatly simplifies the fitting process, it introduces unwanted systematic errors that grow with time (and with different observing locations). 
Despite these differences, the interpretation of the ring geometry remains largely unchanged from that of previous works. 
%To fully understand whether Q1R is affected by the 6:1 Weywot MMR or 6:2 Quaoar SOR, the radius of the ring must be known to within 10s of km. 

\trackchange{Our fitting here uses all reported observational data (ranging from 2019-2024) and assumes a constant ring pole. The reported difference in ring geometry from our work and the literature does not imply that the ring pole is changing in time, but rather that systematic errors are affecting past determinations of the ring pole. }

Our model can also investigate the mirror ring pole solution identified in \citet{morgado2023dense}. We confirm that the mirror provides a far worse fit to the data, allowing us to exclude it at 3.6$\sigma$ confidence. This further confirms that Weywot's orbit is roughly co-planar with the ring plane. 

To test whether the possible ingress detection of Q2R could be part of a ring inclined to Q1R, we also fit Q2R alone. We find that the four detections of Q2R adequately fit a model of a ring with radius $3210\pm80$ km that is $9.8\pm0.5\degr$ inclined relative to Q1R. \trackchange{Although this is consistent with the ingress detection being part of Q2R, theoretical considerations of inclined rings (which are further discussed in Section \ref{sec:discussion}) cast extreme doubt on this hypothesis. }

\begin{deluxetable}{lcc}
\tablecaption{Quaoar Ring Geometry}
\tablehead{
Parameter            &            & Q1R+Q2R fit
}
\startdata
Q1R radius (km)      & $r_1$      & $4096\pm10$                    \\
Q2R radius (km)      & $r_2$      & $2529\pm12$                    \\
Ring pole RA (deg)    & $\alpha_1$ & $259.5\pm0.2$                 \\
Ring pole declination (deg) & $\delta_1$ & $55.0\pm0.2$                  \\
\hline
Opening angle on 2024-08-28 (deg) & & $18.1\pm0.2$ \\
\enddata
\tablecomments{Column titled Q1R+Q2R fit shows the simultaneous ring fit, without the possible ingress Q2R detection. All angles are in J2000 equatorial coordinates. \trackchange{Opening angle is that as viewed from JWST, which provides a slightly different viewing angle as compared to geocentric observations.}}
\label{tab:fits}
\end{deluxetable}

\begin{figure}
    \centering\includegraphics[width=0.49\textwidth]{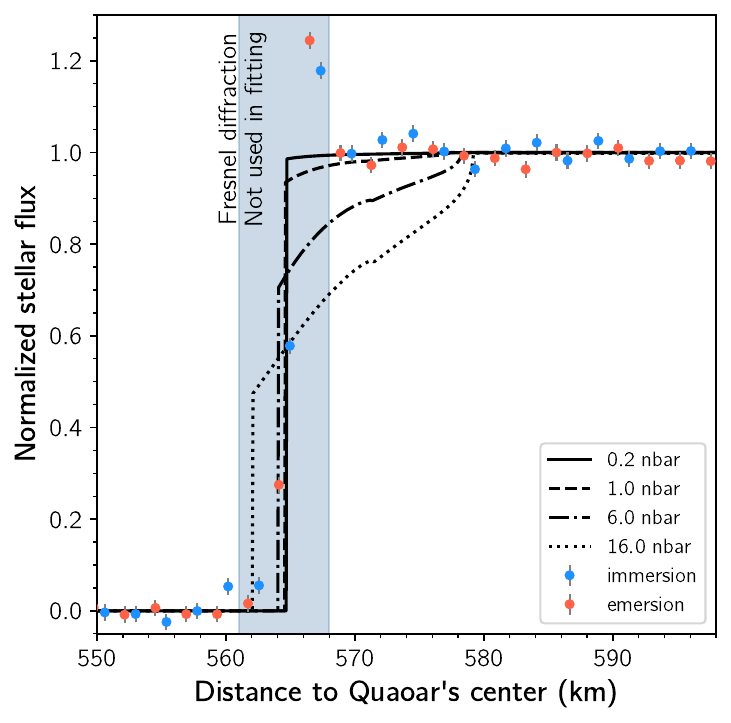}
    \caption{The normalized light curve overlaid with synthetic light curves based on pure ${\rm CH_4}$ atmospheric models. The immersion and emersion profiles are shown as blue and red circles with error bars, respectively.
    Solid and dashed lines represent synthetic light curves corresponding to CH$_4$ surface pressures $p_{\rm surf}$ of \trackchange{0.2} nbar (1$\sigma$ upper limit) and \trackchange{1} nbar (3$\sigma$ upper limit).
    Dash-dotted and dotted lines indicate models with surface pressures of 6 nbar and 16 nbar, respectively, which correspond to the 1$\sigma$ and 3$\sigma$ upper limits derived from previous occultation studies \citet{arimatsu2019new}.}
    \label{fig:atmosphere}
\end{figure}

\section{Atmospheric Limits}
\label{sec:atmosphere}
%This is just some placeholder stuff based on Ko's email.

As seen in Figure \ref{fig:lc}, our observations exhibit remarkable temporal resolution and photometric precision, showing prominent diffraction effects near Quaoar immersion and emersion. The mere presence of these diffraction effects themselves demonstrate that Quaoar has little global atmosphere. However, to place quantitative limits on Quaoar's putative atmosphere, we have modeled atmospheric refraction during the immersion and emersion phases of the solid body occultation. Past studies of Quaoar's atmosphere have not considered diffraction effects, which greatly increase the computational cost of modeling occultation light curves, especially when considering realistic atmospheric pressure/temperature profiles. Instead, we focus on the regions of the light curve outside of occultation where diffraction effects are minimal and where atmospheric refraction can still be detected.

Based on the methods described in \citet{arimatsu2019new}, we fitted a model light curve assuming a pure CH$_4$ atmosphere to the immersion and emersion data between 568 km to 598 km from the predicted center of Quaoar's shadow. 
In this model, the vertical number density profile of CH$_4$ molecules is calculated by integrating the hydrostatic equation for an ideal gas from the surface upwards, assuming a surface pressure of $p_{\rm surf}$.
For the vertical temperature profile, we adopt a Pluto-like structure, where heating of CH$_4$ molecules due to their absorption of solar near-infrared radiation increases the temperature $T$ from $T = 44$ K at the surface to $T = 102$ K in the stratosphere 
(at an altitude of $\sim 10$ km).
We used a density of 1.72 g cm$^{-3}$ for Quaoar, as reported by \citet{kiss2024visible}, and assumed a radius of 550 km, corresponding to a surface gravity of $0.26 \ {\rm m\, s^{-2}}$.

Figure~\ref{fig:atmosphere} presents the synthetic light curves of the models overlaid with the observed immersion and emersion profiles.
The results of the fit indicate that the $1\sigma$ and $3\sigma$ upper limits on the surface pressure $p_{\rm surf}$ of a global atmosphere are \trackchange{0.2} nbar and \trackchange{1} nbar, respectively.

As suggested in other works, Quaoar could also host a local atmosphere generated near the subsolar point \citep{braga2013size,arimatsu2019new}. Instead of creating a dome, thermal inertia and Quaoar's rotation may create an extended local atmosphere along the subsolar latitude that may extend to Quaoar's limb \citep[see Figure S4 of][]{ortiz2012albedo}. Fortunately, our occultation chord closely probes the subsolar latitude on immersion and emersion. Although we do not perform explicit modeling of any local atmosphere, the upper limits from our global atmosphere modeling show that any local atmosphere present does not appear to extend to Quaoar's limb.

\section{Discussion}
\label{sec:discussion}

\subsection{Possibilities of a Third Ring}
The simplest explanation for the unexpected flux drop\trackchange{---aside from it being a statistical outlier---}is the presence of a third ring around Quaoar at a radius of $\sim2300$ km. This ring would be extremely thin ($<2$ km), but with greater opacity than our detections of Q1R or Q2R, although we only put an upper/lower limit on the width/opacity of this putative structure. Just like Q1R and Q2R, this ring could have substantial azimuthal variations which prevented detection during egress. The non-detection of a similar flux drop in the F322W2 filter may imply that this structure is made up of smaller particles, which absorb less of the occulted star's light. We leave full radiative transfer modeling of this scenario to future work, if this putative ring is confirmed to exist.

In an attempt to confirm the presence of a new ring, we searched for similar structures in various occultation light curves in the literature. 
%Comparing light curves is difficult given the different chord geometry between observers and events. To allow easier comparison, we transform the light curve (which is a function of time) into a function of radial distance from Quaoar (assuming a specific ring geometry), eliminating the effects of chord geometry during the comparison. This process allows easy vetting of candidate ring structures even within a single light curve. 
We find no evidence for a similar structure in high quality occultation chords from Gemini, CFHT, or GTC \citep{pereira2023two, morgado2023dense}. However, if the structure has azimuthal variations---like Q1R and Q2R---it may have been able to avoid detection. \trackchange{The non-detection in the LWC, as well as no confirming observations from other occultation events, strongly suggest that this putative detection is, in reality, just an unluckily positioned 4$\sigma$ outlier. In either case,} confirming or rejecting the existence of any putative additional structures will require more occultation observations with large telescopes using high-cadence imagers. Events of particular interest are those with bright stars and/or slow on-sky velocity, which allow for higher signal-to-noise and better spatial resolution. 

\subsection{A Non-Coplanar/Eccentric Q2R}
\trackchange{Presuming again the unlikely case that the ingress detection represents a real structure,} one possibility is that the detection is Q2R. This would imply that Q2R is either non-coplanar with Q1R and/or is not circular. Past works only assumed that the ring had the same pole orientation as Q1R due to lack of detections, but with two possible new detections, we can systematically test this hypothesis. \trackchange{Using the ring modeling techniques described in Section \ref{sec:model},} we find that Q2R may be compatible with a $\sim3200$ km circular ring inclined by $9.8\pm0.5\degr$ to Q1R. 

Maintaining an inclined ring is theoretically difficult. Differential precession of the nodes of orbiting particles significantly increase collisional activity, which tends to damp the inclination of the ring to the local Laplace plane \citep{tiscareno2014planetary}. The Laplace plane is the (local) plane around which orbits of test particles precess, the orientation of which is determined by the surrounding dynamical environment. For example, in the presence of an oblate gravitational field (like that of Quaoar), a ring will tend to damp to the equatorial plane. In the presence of a satellite, however, ring particles can instead damp to the satellite orbit plane, if the satellite perturbations are strong enough \citep{marzari2020ring}. It may be that the dynamical environments of Q1R and Q2R are radically different, allowing them to damp to different orbital planes. For example, Q1R may be strongly perturbed by Weywot's 6:1 MMR, while Q2R may be coupled more strongly to Quaoar's oblate gravitational field. 

As a slight modification of this hypothesis, Q2R could also have substantial eccentricity. The maximum radial variation in an orbit due to eccentricity is $2 a e$, where $a$ is the ring semi-major axis and $e$ is the eccentricity. The radial variation from observed Q2R occultations provides a lower limit on the eccentricity,  $e_{\mathrm{Q2R}} \gtrsim 0.04$, an order of magnitude larger than other eccentric rings \citep[e.g.][]{french1991dynamics}. Again, theoretical considerations show this may be difficult, although some slightly eccentric rings are maintained by viscous forces \citep{chiang2000apse}, self-gravity \citep{goldreich1979towards}, or shepherding from nearby small satellites \citep{melita2020apse}. We also note that both possibilities could play a role in explaining Q2R. The ring could be both eccentric and inclined. We would expect that the estimates/limits placed ($e\gtrsim0.04$, $i \approx 10\degr$) would be somewhat decreased if both inclination and eccentricity were present. 

\trackchange{Given the theoretical difficulties associated with an inclined or eccentric Q2R, this again suggests that the most parsimonious explanation for the ingress detection is a poorly placed outlier. This can be confirmed with additional occultation observations of Q2R. Any eccentricity/inclination will be easily detectable in future occultations, with only a single detection needed to rule out these hypotheses. Particular effort should be focused on occultations near quadrature, which can have North-South geometry, allowing us to probe the rings at unexplored azimuths.}

\subsection{Azimuthal Variations in the Rings}
Combined with our two detections, Q1R has now been detected in 19 different occultation chords over 6 years \trackchanget{\citep{morgado2023dense,pereira2023two}}. The occultation chords show strong evidence for drastic azimuthal variations, a result which is also seen in our detections. Our occultation chords failed to detect the densest portion of Q1R, which can have apparent opacity of $\gtrsim$0.5. Non-detections still provide useful information about the azimuthal arc length of the dense portion of the ring. Treating the (non)detections of dense arcs as a series of pass-fail trials, with two successes from 21 trials (where a statistically independent chord has sufficient signal-to-noise to detect the dense portion of Q1R; \citealt{pereira2023two}), we find that the total arc length of the dense portion of the rings is between 17-66$\degr$ at 70\% confidence, or 10-104$\degr$ at 95\% confidence. If Q1R has multiple similarly deep arcs (like Neptune's Adams ring), this constraint is on the sum of the arc lengths.

Discarding the possibility of an inclined and/or eccentric Q2R, our observations show that Q2R must have substantial azimuthal variability due to the non-detection of Q2R on ingress. Using the same statistical argument for Q1R is difficult as no reanalysis of light curves from before Q2R's discovery has been publicly released. So just using the discovery of Q2R along with our work (the only light curves which are known to be able to detect Q2R), we find the ring has an arc length of 177-325$\degr$ (70\% confidence) or 108-344$\degr$ (95\% confidence). Clearly, these four detections are not sufficient to say much about the variability or azimuthal structure of Q2R, other than confirming it is variable. 

\subsection{The Dynamical Environment of the Rings}
Detailed comparisons of Q1R's geometry with Weywot's orbit allow us to determine Weywot's contribution to the ring's dynamics. Using updated orbit fits from recent Hubble Space Telescope (HST) observations, Weywot's semi-major axis has been tightly constrained to be $a = 13300\pm40$ km (Proudfoot et al., in prep.). This places Weywot's 6:1 resonance at $4028\pm12$ km, inside of Q1R at $4097\pm10$ km. However, considering the ring's width of up to 300 km \citep{morgado2023dense}, the 6:1 may strongly affect the inner edge of Q1R. 

The oblateness of Quaoar splits the 6:1 into various sub-resonances, whose exact radial locations can vary based on the apsidal and nodal precession rates of Weywot \citep{1999ssd..book.....M}. These precession rates are currently unconstrained, but continued precise observations of Weywot's position may be able to measure them. Until those measurements, the dynamics of Q1R as they relate to Weywot's 6:1 MMR will remain unclear. In future work, we plan to directly study the (sub)resonance locations, widths, and strengths, which should shed further light on the dynamics of Q1R.

In addition to MMR dynamics with Weywot, Q1R is close to Quaoar's 3:1 SOR, which lies at $4217\pm26$ km (Proudfoot et al., in prep., \citealt{kiss2024visible}). The 3:1 SOR seems to be important for small-body ring dynamics, with Haumea's rings at (or near) its 3:1 SOR \citep{ortiz2017size} and Chariklo's plausibly near its 3:1 SOR \citep{sicardy2019ring}. Based on visible and thermal light curve measurements, Quaoar may be substantially elongated ($\epsilon = \frac{a^2-b^2}{2R^2} \sim0.2$) making the 3:1 a source of strong dynamical interactions \citep{kiss2024visible}. Again considering the width of Q1R, the 3:1 SOR seems likely to affect the outer edge of the ring. Depending on the width and strength of the 3:1 SOR, it may be possible that some parts of Q1R lie outside of the 3:1 SOR. Unfortunately, the understanding of how rings interact with the 3:1 SOR of their parent bodies is still in its infancy. Simplified theoretical models provide some guidance on the dynamics \citep[e.g.][]{sicardy2019ring,sicardy2020resonances,sicardy2020dynamics}, but models that incorporate other important physical effects (e.g., particle collisions, self-gravity, higher-order gravitational harmonics) have yet to be fully explored. 

%SORs around small bodies can be caused by either a triaxial shape or mass anomalies of topographic features. The strength of the 3:1 SOR, as caused by a triaxial body's shape, is proportional to elongation, $\epsilon$, where $\epsilon = \frac{a^2-b^2}{2R^2}$ and $a, b$ are the semi-axes of the long and intermediate axes, and $R$ is the body's reference radius. For mass anomalies, the SOR strength are proportional to $\mu\sim(z/2R)^3$, where $z$ is the size of a topographic feature (like a crator or a mountain) and $R$ is the radius of the body \citep{sicardy2020dynamics}. 

%To date, little is known about Quaoar's three-dimensional shape. Occultation chords show Quaoar has substantial oblateness \citep{braga2013size,pereira2023two}, with the oblateness varying between events. This implies Quaoar has a triaxial shape, but a full three-dimensional model from precise occultation chords has not yet been tested. Recent analysis of thermal observations of Quaoar suggest a triaxial shape with $\epsilon\sim0.2$, but require a substantial departure from a hydrostatic equilibrium figure \citep{kiss2024visible}. Topography on Quaoar has yet to be seen, but given Quaoar's density and the typical material strength of icy bodies, Quaoar is unlikely to be able to support any topographic features $z\gtrsim10$ km \citep{johnson1973topography}. In comparison, terrain on Pluto's encounter hemisphere is $\lesssim6$km \citep{schenk2018basins}. This constrains $\mu\lesssim 10^{-6}$, implying that, at least for the 3:1 resonance, Quaoar's shape plays a dominant role in the spin-orbit dynamics. 

With the inner edge of Q1R bounded by Weywot's 6:1 MMR (and its various sub-resonances) and its outer edge bounded by Quaoar's 3:1 SOR, it is clear that Q1R's dynamical context is extremely complex. The collisional activity driven by the dynamical environment, combined with the argument that collisions are more elastic at the temperatures of the Kuiper belt \citep{morgado2023dense}, may be enough to prevent the accretion of Q1R. Collisions, however, would cause viscous spreading of the rings. Confinement may be enabled by both bounding resonances. Now, with precise determinations of the ring geometry, constraints on Quaoar's triaxial shape \citep{kiss2024visible}, and better measurements of Weywot's orbit (Proudfoot et al., in prep.), full dynamical simulations should be able to shed light on the confinement and stability of Q1R. 

The presence of the inner, fainter ring, Q2R, remains an interesting mystery. Unlike Q1R, Q2R is not near any significant MMRs with Weywot. The 7:5 SOR is located at $2537\pm15$ km, very near Q2R at $2529\pm12$ km. While Q2R is close to the 7:5 SOR, the 7:5 SOR is fairly weak. Although fourth-order like the 3:1 SOR, the strength of SORs caused by triaxial shapes scales with $\epsilon^{|m/2|}$, where $m=-10$ for the 7:5 and $m=-2$ for the 3:1 \citep{sicardy2020resonances}. With Quaoar's elongation ($\epsilon\sim0.2$), the 7:5 SOR would be $10^{-3}$ times the strength of the 3:1 SOR. In addition to the relatively weak SOR strength, extensive simulations of the clearing of particles around triaxial bodies show that particles within the 2:1 SOR ($\sim$2650 km) are pushed outwards on $\sim$decade-long timescales \citep{sicardy2019ring}. Both of these factors cast doubt on the role of the 7:5 SOR---and spin-orbit interactions as a whole---on the confinement of Q2R. 

Q2R is also reasonably close to Weywot's 12:1 MMR at $2537\pm8$ km. The 12:1 is made up of 12 sub-resonances (where $k$ takes the value 0-11), with 1 co-rotation type resonance and 11 Lindblad resonances. Each sub-resonance has strengths proportional to $e^ke_w^{11-k}$, where $e, e_w$ are a ring particles and Weywot's eccentricity, respectively. Although this would traditionally predict very weak resonances, $n$:1 MMRs can actually be unexpectedly strong if the perturber is on a very eccentric orbit. Since Weywot is on a nearly circular orbit, with $e<0.02$ (Proudfoot et al., in prep.), we expect that the 12:1 MMR likely exerts little influence on Q2R. 

\subsection{Possible Shepherd Moons}
Instead, Q2R may be controlled by one or more shepherd moons. Shepherd moon(s) have been suggested as a mechanism for the confinement of Chariklo's rings, with various works suggesting at least one moon is required \citep[][]{braga2014ring,melita2020apse,winter2023stability}. In addition to confinement, recent numerical simulations of Chariklo's rings (which are possibly outside of Chariklo's Roche limit) have shown that a single shepherd can simultaneously confine rings and protect against accretion \citep{sickafoose2024chariklo}. A delicate balance between accretion and destruction can occur, where forming moonlets in the wakes of shepherds are quickly sheared apart. Although Quaoar is an order of magnitude larger than Chariklo, similar mechanisms could be at play for Q2R. 

At present, detection of shepherd satellites around Quaoar remains difficult. Shepherd satellites with diameters of 1-10 km would have apparent magnitudes of $V\gtrsim27$ with angular separation from Quaoar of $\lesssim$0.15 arcsec. In comparison, even Keck and HST struggle to detect Weywot ($V\approx24.7$) within $\sim$0.2 arcsec of Quaoar \citep{fraser2010quaoar}. In fact, it is entirely possible that Weywot-sized satellites could possibly still remain undetected interior to Weywot's orbit. 

One possible way to detect (or place upper limits on) shepherd satellites is through observations of additional stellar occultations. 
Observations of recent stellar occultation by Huya have proved useful in detecting and characterizing the orbit of Huya's satellite, which is at the detection limit of HST and Keck \citep{rommel2025stellar}.
If campaigns of dozens (or even hundreds, see \citealt{rommel2023ms4}) of observers can simultaneously probe Quaoar and its surroundings, any small satellites in the vicinity of Quaoar's ring may be discovered. 
Even with enough non-detections, useful upper limits may be able to be derived on the presence of any undetected satellites.
Alongside observations, future theoretical and numerical simulations of Quaoar's rings should focus on the possibility of inner shepherd moons.

In addition to direct searches for shepherds, it may be possible to infer their presence by tracking the location of the Q1R arc(s). With many detections across various epochs the mean motion of the arc(s) can be measured, providing insight into the source of their confinement. Similar studies have proved valuable for understanding the confinement of arcs within Neptune's Adams ring \citep[e.g.,][]{nicholson1995stellar,Pater_Renner_Showalter_Sicardy_2018}. 

\subsection{Atmosphere}

The upper limits on the surface pressure of a potential atmosphere obtained in this study are significantly lower than those derived from previous occultation observations, which were 6 nbar and 16 nbar for the $1\sigma$ and $3\sigma$ upper limits, respectively \citep{arimatsu2019new}.
The non-detection of an atmosphere around Quaoar, while unsurprising, has profound implications. Recent NIR spectroscopy from JWST's NIRSpec instrument have revealed that light hydrocarbons like CH$_4$ and C$_2$H$_6$ are present on the surface of Quaoar. Irradiation of these molecules is rapid and their surface inventories would be quickly depleted, suggesting an ongoing resupply to the surface \citep{emery2024tale}. For Quaoar, the non-detection of an atmosphere shows seasonal volatile migration---like that on Pluto \citep{young2021pluto}---is an unlikely source of this resupply. Further, its relatively circular orbit ensures a perihelion atmosphere is unlikely to form. 

One plausible source of ongoing CH$_4$ supply is from internal geochemical processing, such as has been suggested for Makemake and Eris \citep{glein2024moderate,grundy2024measurement}. Quaoar is presumably large enough that some portions of its interior could have undergone similar processing. If this is the case, stochastic phases of outgassing or geological activity could periodically refresh Quaoar's surface CH$_4$ while remaining consistent with our non-detection of an atmosphere. This process could be connected to the generation of Quaoar's ring system. Given the rings' questionable long-term stability, periodic surface activity may play a role in refreshing the ring system.

Another plausible source of surface CH$_4$ is primordial CH$_4$ delivered by impact. \trackchange{Impacts could also excavate fresh CH$_4$ stored beneath the thin crust of radiation processed hydrocarbons.} The viability of this route could be further assessed based on predicted impact rates and CH$_4$ irradiation lifetimes.

\section{Conclusion}
\label{sec:conclusion}
In this work, we present JWST NIRCam observations of a stellar occultation by Quaoar and use them to probe Quaoar's rings and putative atmosphere. We confidently detect Quaoar's two known rings, confirming the \trackchange{continued presence of both} rings. \trackchange{Based on an unexpected flux drop near the location of Q2R along with a non-detection of Q2R on ingress, we hypothesize that Q2R could be eccentric/non-coplanar (with the outer ring) or that another ring may be present 200 km inside of Q2R. Alternatively, the unexpected drop could be a poorly-timed low-probability outlier. Theoretical considerations strongly favor the random outlier explanation, however, future observations can more definitively confirm this.} In either case, our observations show that Q2R has substantial variations in width and/or depth, although further detections of Q2R are necessary to fully place into context our observations. 

Combining the various detections of Q1R and Q2R found in the literature with our work, we are able to fully model the size and orientation of the ring system. Our results are inconsistent with previous work \citep{morgado2023dense, pereira2023two} at high confidence ($>5\sigma$)\trackchange{, likely due to systematic factors rather than changing ring orientation}. The improved constraints on the radii of the two rings shows that the inner edge of Q1R is likely affected by Weywot's 6:1 MMR. Further dynamical modeling is needed to understand the structure of the Weywot 6:1 MMR, with special attention given to resonance splitting induced by Quaoar's putative oblateness. Likewise, evaluation of whether Quaoar's 3:1 and 7:5 SORs play a role in the confinement of Q1R and Q2R requires further dynamical modeling. Based on the possible inability of these SORs to confine the rings---especially in the case of the 7:5---we suggest that one or more shepherd moons may be present, which could be detectable by continued observations of stellar occultations. 

Analysis of our observations near immersion/emersion around Quaoar suggest that Quaoar possesses no pure CH$_4$ global atmosphere with surface pressures greater than \trackchange{1} nbar (3$\sigma$). This ultra-low upper limit clearly demonstrates that recently detected light hydrocarbons on Quaoar's surface are not replenished by ongoing atmospheric processes. Instead, we suggest periodic geologic activity \trackchange{or impacts} could stochastically replenish Quaoar's surface inventory of light hydrocarbons, which may contribute to the production or maintenance of the ring system. 

\section*{Acknowledgments}
We thank John Stansberry, Darin Ragozzine, and Jose Luis Ortiz for helpful discussions that helped improve our manuscript. We also thank two anonymous reviewers who helped make this work substantially clearer. We acknowledge the Lucky Star collaboration for their work to continually update the ephemerides of TNOs.

This work is based on observations made with the NASA/ESA/CSA James Webb Space Telescope. The data were obtained from the Mikulski Archive for Space Telescopes at the Space Telescope Science Institute, which is operated by the Association of Universities for Research in Astronomy, Inc., under NASA contract NAS 5-03127 for JWST. These observations are associated with program \#6780. Support for program \#6780 was provided through a grant from the STScI under NASA contract NAS5-03127.

B.P. and F.R. gratefully acknowledge the support of the University of Central Florida Preeminent Postdoctoral Program (P3). B.H. acknowledges support from NASA SSO 19-SSO19\_2-0061.

\bibliographystyle{apj}
\bibliography{all}

\end{document}